# Testing Lorentz symmetry of special relativity by means of the Virgo or Ligo set-up, through the differential measure of the two orthogonal beams time-of-flight

G. SARDIN (*)

*Applied Physics Department, University of Barcelona - Barcelona, Spain*

**Abstract.** - A novel experiment to test special relativity via Lorentz symmetry has become factible thanks to three recent technological achievements: huge Michelson-like set-up with arms 3 km long (Virgo) and 4 km (Ligo) with beam paths respectively reaching 120 km and 200 km through multiple reflections, ultrashort laser pulses of $10^{-15}$ s and ultrafast detectors of $10^{-12}$ s resolution. The alliance of these three elements would allow checking the equality of the time-of-flight of the two orthogonal beams with a resolution high enough to allow prospecting in a novel way the equivalency of inertial system postulated in special relativity. In effect, for a beam path length of 120 or 200 km and a net drift velocity of earth of 370 km/s relative to the cosmic microwave background (CMB), a classical analysis predicts a time-of-flight difference of the order of $10^{-9}$ s between the two orthogonal beams, while relativity infers them to be equal. So, what is under scrutiny is the exhaustivity of the electromagnetic equivalency of inertial systems. A null time-of-flight difference would strengthen the Lorentz symmetry, while a non null result would bring a threshold to the equivalency of inertial systems and at the same time would provide a tool to define their speed, which should be equal to that relative to the CMB for being congruent.

*Introduction* - An original experiment, exploiting the Virgo (1) or Ligo (2) set-up for testing Special relativity in a still unsound aspect rooted in the direct access to the differential measure of the two orthogonal laser beams time-of-flight, is proposed. In effect, these gravitational wave observatories rely on a huge Michelson-Morley interferometer (3), whose two orthogonal arms measure 3 km and 4 km respectively and whose beam path lengths reach 120 km and 200 km through multiple reflections. Such lengthy paths allow the differential measure of the beams time-of-flight. The new test would not rely in evidencing the phase invariance versus speed of the two orthogonal beams upon recombination, but instead on the direct detection of any time-of-flight difference between them. This type of measurement was unworkable up to now and only the alliance of three modern technological advances, specifically the outstanding features of the Virgo and Ligo set-up, ultrafast laser with femtosecond pulses and ultrafast time-of-flight detectors of picosecond resolution, allow to envision reaching this incisive and crucial experiment.

Let us rough out the conjecture. The principle of Lorentz invariance stipulates that the result of any experiment must be formally independent of the speed of the inertial system on which it is performed. Explicitly, since the first postulate of special relativity states that inertial systems are equivalent, this is to say that the time-of-flight of the two orthogonal beams must be independent of speed otherwise Lorentz symmetry (4) would be infringed. This is precisely the point that the experiment would insure since the interpretation of its result does bear any ambiguity, which is not the case in electromagnetic experiments relying on the wave nature of light such as Michelson-type interferometric experiments (5), as we have pointed out, suggesting that the null result can be interpreted as being due to the influence of the Doppler effect on the wave period (6). This misgiving applies to all experiments relying on the beams wave-length, period or phase, inducing mistrusts on the factual cause for Lorentz invariance settled in special relativity. Therefore, the direct access to the differential measure of the beams time-of-flight would bring a decisive proof on Lorentz symmetry, free of any reliance on the two beams phase and thus of any imponderable influence of the Doppler effect.

So, in order to stay away from any interpretative doubts, the null time difference between the two beams time-of-flight in Michelson-like experiment is aimed to be checked taking advantage of the pertinent modern facilities, which would allow detecting any deviation from conformity with Lorentz symmetry. The classical analysis of the time-of-flight through ray optics predicts a time difference between the two beams: $\Delta t = t_x - t_y = (d / c) (v / c)^2$. Knowing that the earth has a mean net drift velocity relative to the CMB of the order of 370 km/s and for beam paths of 120 or 200 km, a classical viewpoint predicts a time difference of the order of $10^{-9}$ s, time which fits within the ultrafast detectors range of reach (7-15). This time difference corresponds to the case in which one arm is aligned with the direction of the speed vector and represents its maximum value (fig.1), but it is null when oriented along the meridian axis of the two arms due to symmetry (fig.2). According to special relativity the time difference should be independent of any orientation and always null. Since the two beams make a round trip back to the same point of incidence, the differential measure of their time-of-flight can be achieved with a single photodetector.

***Experimental and methodology*** - As already expressed, the experiment addresses to the time-of-flight method instead of the interferometric method. Let us evaluate the time-of-flight difference to be detected by means of such lengthy orthogonal beams path:

$t_x = (d / c) / (1 - v^2 / c^2) \approx (d / c) (1 + v^2 / c^2)$

$$t_y = (d / c) / \text{sqr} (1 - v^2 / c^2) \approx (d / c) (1 + 1/2\, v^2 / c^2)$$

$$\Delta t = t_x - t_y = (d / c) (v / c)^2$$

For a beam path of 120 or 200 km and a speed of the inertial system of 370 km/s:

$$\Delta t = t_x - t_y = (d / c) (v / c)^2 = (120/3\ 10^5) (370/3\ 10^5)^2 \approx 0.6 \times 10^{-9}\ \text{s} = 0.6\ \text{ns}$$

$$\Delta t = t_x - t_y = (d / c) (v / c)^2 = (200/3\ 10^5) (370/3\ 10^5)^2 \approx 1.0 \times 10^{-9}\ \text{s} = 1.0\ \text{ns}$$

Comparatively, in the Michelson-Morley experiment the arm length was of 1 m and the beam path was of 11 m through multiple reflections, therefore the corresponding classical time-of-flight difference between the two orthogonal beams would be:

$$\Delta t = (d / c) (v / c)^2 = (11/3\ 10^8) (370\ 10^3/3\ 10^8)^2 = 4.10^{-14}\ \text{s} = 40\ \text{fs}$$

and thus it is not directly detectable.

The procedure to ascertain the results would mainly rely on three check-up:

(a) To insure that any time-of-flight difference would not be due to a difference of length between arms the measure would be repeated, interchanging the arm aligned in the direction of our galaxy drift motion, using the earth spin to get the change in the arms orientation. If due to a difference of the arms length the time-of-flight difference would not vary, but if not the sign of the difference would be inverted.

(b) Also, another check up would consists on orienting the axis of symmetry of the system in the direction of our galaxy drift (fig.2), the time difference being then expected to be null whatever the system speed. Any time-of-flight difference would thus be to a difference between of the arms length or to an artefact.

(c) Still, an extra assessment implies that, when one arm is oriented in the earth cosmic drift, the speed derived from an eventual time-of-flight difference between the two orthogonal beams should be equal to its drift speed relative to the CMB. Any tangible departure from equality would be incongruent.

*Conclusion* - The proposed experiment would allow to test in a still unsound way the equivalency of inertial systems, and possibly to ratify Lorentz symmetry. So, what would be under scrutiny is the exhaustivity of the electromagnetic equivalency of inertial systems. An invariantly null time-of-flight difference between the two orthogonal beams would strengthen the first postulate while a non-null result would bring a threshold to observer Lorentz invariance, and at the same time would allow to get a new way to define the drift speed of earth through the cosmos. Both speeds, that relative to the CMB and any eventual one deriving from a non null result should be equal. Any deviation from equality would jeopardise the reliability of the result. On the contrary, a systematically null result would extend the equivalency of inertial systems and still strengthen Lorentz symmetry postulated in special relativity. It would thus be a misuse not to take advantage of the extraordinary possibilities offered by the Virgo and Ligo huge set-up to extend their usefulness beyond a circumscribed endeavour to detect gravitational waves. Besides, let stress that the proposed time-of-flight experiment represents a less challenging aim.

# Figures

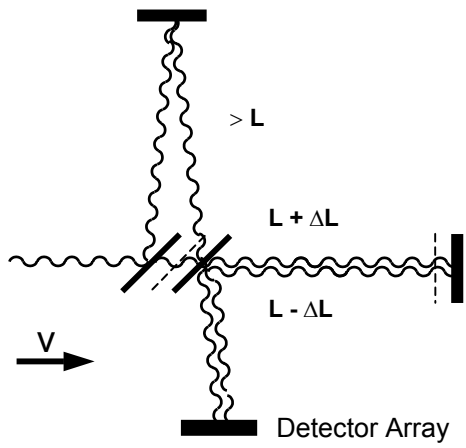

**Fig. 1:** Optical device with one arm oriented in the direction of motion

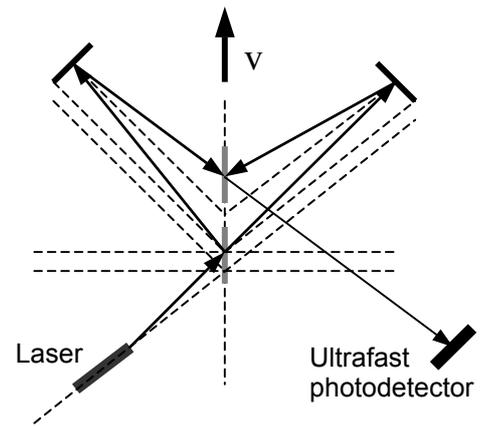

**Fig. 2:** Optical device symmetrically oriented in the direction of motion